\documentclass[numberedheadings]{aipproc}
\usepackage{epsfig}
\usepackage[]{natbib}
\usepackage{aas_macros}     
\usepackage{amsfonts}
\layoutstyle{8x11single}

\begin{document}
\title{How BAO measurements can fail to detect quintessence}

\author{E.~Jennings}{
  address={Institute for Computational Cosmology, Department of Physics, University of Durham, South Road, Durham, DH1 3LE, U.K.}
	,altaddress={Institute for Particle Physics Phenomenology, Department of Physics, University of Durham, South Road, Durham, DH1 3LE, U.K.}
}

\author{ C.~M.~Baugh}{
  address={Institute for Computational Cosmology, Department of Physics, University of Durham, South Road, Durham, DH1 3LE, U.K.}
}

\author{R.~E.~Angulo}{
  address={ Max Planck Intitute fur Astrophysik, D-85741 Garching, Germany.}
}

\author{S.~Pascoli}{
  address={Institute for Particle Physics Phenomenology, Department of Physics, University of Durham, South Road, Durham, DH1 3LE, U.K.}
}

\date{}

\begin{abstract}
We model the nonlinear growth of cosmic structure in different dark energy
 models, using large volume N-body simulations.
 We consider a range of quintessence models
 which feature both rapidly and slowly varying dark energy equations of state, 
and compare the growth of structure to that in a universe
 with a cosmological constant.
 We use a four parameter equation of state 
for the dark energy which accurately reproduces the 
quintessence dynamics over a wide range of redshifts.
The adoption of a quintessence model changes the expansion history of the universe, the form of 
the linear theory power spectrum and can alter key observables, such as the horizon scale and the 
distance to last scattering.  
The difference in structure formation can be explained to first order by the difference in growth factor at a given epoch; 
this scaling also accounts for the nonlinear growth at the 15\% level.
We find that quintessence  models which feature late $(z<2)$, rapid transitions towards $w=-1$ in the equation of state,
 can have identical baryonic acoustic oscillation (BAO) peak positions
to those in $\Lambda$CDM, despite being 
 very different from $\Lambda$CDM both today and at high redshifts $(z \sim 1000)$. 
We find that a second class of models which feature non-negligible amounts of dark energy at early times  
 cannot be distinguished 
from $\Lambda$CDM using measurements of the mass function or the BAO.
These results highlight the need to accurately model quintessence dark energy in N-body simulations 
when testing cosmological probes of dynamical dark energy. 

\end{abstract}

\maketitle

\section{Introduction}

Determining whether or not the dark energy responsible for the accelerating expansion of the Universe evolves with time remains a key goal of physical cosmology.
 This will tell us if the dark energy is indeed a cosmological constant or has a 
dynamical form as in quintessence models.
The nature of the dark energy determines the expansion history of the Universe and hence the rate at which cosmological perturbations grow. In this paper we investigate the 
influence of quintessence dark energy on the nonlinear stages of structure formation using a suite of N-body simulations.
In quintessence models, the cosmological constant  is replaced by an extremely light scalar field
which evolves slowly \citep{Ratra:1987rm,1988NuPhB.302..668W,1998PhRvL..80.1582C,Ferreira:1997hj}. 
 The form of the 
scalar field potential  determines the trajectory of the equation of state, $w(z)=P/\rho$, as it evolves in time.
Hence, different quintessence dark energy models have different dark energy densities as a function of time, $\Omega_{\tiny \mbox{DE}}(z)$.
This implies a different growth history for dark matter perturbations from that expected in $\Lambda$CDM. 

Cosmological N-body simulations are the theorist's tool of choice for modelling the final stages of perturbation collapse.
The overwhelming majority of simulations have used the concordance $\Lambda$CDM
cosmology. Here we simulate different dark energy models and study their 
observational signatures.
A small number of papers have used N-body simulations to test 
scalar field cosmologies 
 \citep{ Ma:1999dwa,Klypin:2003ug, Francis:2008md,Grossi:2008xh,2009arXiv0903.5490A}. 
Rather than explicitly solving for different potentials, it is standard practice to modify the Friedmann equation using a form for the dark energy
equation of state, $w(z)$.
Previous work has
used a variety of 
parametrizations for  $w(z)$, the most common being the two parameter  equation, $w = w_0 + (1-a)w_a$
 \citep{Chevallier:2000qy, Linder:2002et} or the empirical three parameter equation proposed by \citet{Wetterich:2004pv} for the so-called early dark energy models. 
The disadvantage of using a 1 or 2 variable parametrization for $w$ is that it cannot accurately reproduce the dynamics of 
a quintessence model over a wide range of redshifts.
 Instead, we take advantage
of a parametrization for $w(z)$ which can describe a variety of different models.
In this work we use a four parameter dark energy equation of state  which can accurately reproduce the original $w(z)$
for a variety of dark energy models to better than 5\% for redshifts $z<10^3$ \citep{Corasaniti:2002vg}.

In Section 2 we discuss quintessence models and the parametrization we use for the dark energy equation of state. 
 We also outline the expected impact of different dark energy models on structure formation. In Section 3 we give the  details 
of our N-body simulations. 
The main power spectrum results are presented in Section \ref{wmap}.  
In Section \ref{bao} we discuss the appearance of the baryonic acoustic oscillations in the matter power spectrum. Finally, in Section 5 we present our conclusions.
Further details can be found in \citet{Jennings}.

\section[]{Quintessence Models of Dark Energy \label{QUIN}}

Here we briefly review some general features of quintessence models; more detailed descriptions can be found, for example,  in 
 \citet{Ratra:1987rm,Ferreira:1997hj,Copeland:2006wr} and \cite{Linder:2007wa}.
The dynamical quintessence field is a slowly evolving component with negative pressure. 
The Hubble parameter for dynamical dark energy 
in a flat universe is given by
\begin{equation} 
\frac{H^2(z)}{H_0^2}=\left( \Omega_{\rm m} \,(1+z)^{3} + (1-\Omega_{\rm m}) e^{3\int_0^z 
d ln (1+z') \, [1+w(z')]}\right),
\end{equation}
where $H_0$ and $\Omega_{\rm m} = \rho_{\rm m}/\rho_{{\tiny \mbox{crit}}}$ are the values of the  Hubble parameter 
and dimensionless matter density, respectively, at redshift $z =0 $ and $\rho_{{\tiny \mbox{crit}}} = 3H_0^2/(8\pi G)$ is the critical density.
The dark energy equation of state is expressed as the ratio 
of the dark energy 
pressure to its energy density, denoted as $w = \textrm{P}/\rho$.
Once a standard kinetic term is assumed in the quintessence model,  
it is the choice of potential
 which determines 
$w$ as
\begin{equation}
\label{wgeneral}
w = \frac{\dot{\varphi}^2/2 - V(\varphi)}{\dot{\varphi}^2/2 + V(\varphi)}\,.
\end{equation}
We consider  six quintessence models which  feature both rapidly and slowly varying equations of state (see \citet{Jennings} for more details).
Some of these models display significant levels of dark energy
at high redshifts in contrast to a $\Lambda$CDM cosmology. As the AS \citep{Albrecht:1999rm}, CNR \citep{Copeland:2000vh}, 2EXP \citep{2000PhRvD..61l7301B} and SUGRA
 \citep{Brax:1999gp} models have non-negligible
dark energy at early times, all of these could be classed as \lq early dark energy\rq \, models.
In Section 4 we will investigate if  quintessence models which feature an early or late transition in their equation of state, and in their dark
energy density, can be distinguished from $\Lambda$CDM by examining  the growth of large scale structure.

\begin{figure}
{\epsfxsize=7.truecm
\epsfbox[101 359  469 718]{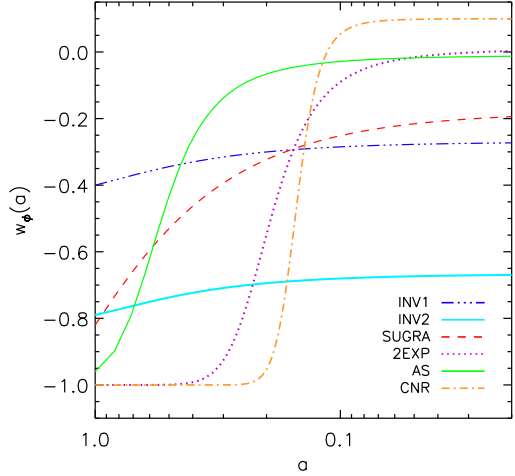}}
\caption{
The dark energy equation of state as a function of expansion factor, 
$w(a)$, for six quintessence models.
The parametrization used for $w(a)$ and the parameter values which specify each model are  given in \citet{Jennings}. 
Note the left hand side of the x-axis is the present day.} \label{w}
\end{figure}

\subsection{Parametrization of $w$}

Given the wide range of quintessence models in the literature
 it would be a great advantage, 
when testing these models, to obtain one model independent equation which could describe the evolution of  the dark energy equation of state
without having to specify the potential $V(\varphi)$ directly. 
We  employ the  parametrization for $w$ proposed by \citet{Corasaniti:2002vg}, which is a generalisation of the method used by \citet{Bassett:2002qu} for fitting dark energy models 
with rapid late time transitions.
We use the 
shorter version of this parametrization for $w$ which depends on four variables, see \citet{Corasaniti:2002vg}, 
which is relevant as our simulations begin in the matter dominated era. 
\citet{Corasaniti:2002vg} showed that this four parameter fit gives an excellent match to the exact equation of state.
The parametrization for the dark energy equation of state  is plotted in Fig. \ref{w} for the various quintessence models used in this paper.

The adoption of a 4 variable parametrization is essential to accurately model the expansion history over the full range of redshifts probed by the simulations.
Using a 1 or 2 parameter equation of state whose application is limited to low redshift measurements 
restricts the analysis  of the properties of dark energy and cannot make use of high redshift measurements such as the CMB. 
\citet{Bassett:2004wz} analysed how accurately various parametrizations 
could reproduce the dynamics of quintessence models. They found that parametrizations based on an 
expansion to first order in $z$ or $\mbox{log} \,z$ showed errors of $\sim 10\%$ at $z=1$.
A general prescription for $w(z)$ containing more parameters than a simple 1 or 2 variable equation can accurately describe both 
slowly and rapidly varying equations of state \citep{Bassett:2004wz}. For example, the parametrization provided by \citet{Corasaniti:2002vg} can accurately mimic the exact time behaviour of $w(z)$
to $<5\%$ for $z<10^3$  using a 4 parameter equation of state and to $<9\%$ for $z<10^5$ with a 6 parameter equation.

\begin{figure}
{\epsfxsize=7.truecm
\epsfbox[96 363 453 701]{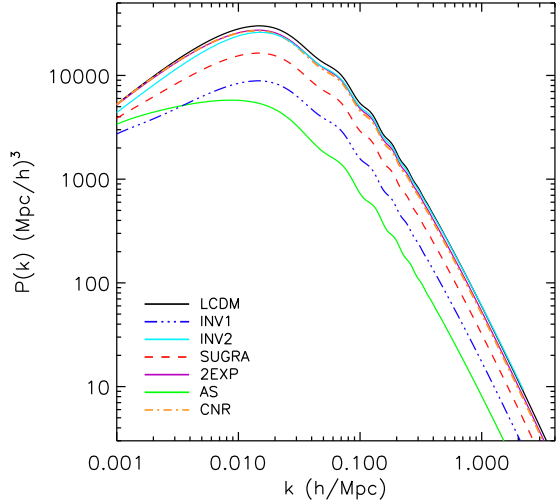}}
 \caption{Linear theory power spectra at $z= 0$ for dynamical dark energy
quintessence models and $\Lambda$CDM.
In this plot, the spectra are normalised to CMB fluctuations (on smaller wavenumbers than are included in the plot).
}
 \label{rawpk}
\end{figure}

\subsection{The expected impact of dark energy on structure formation \label {2.3}}

The growth of structure is sensitive to the amount of dark energy, as this changes the rate of expansion of the Universe.
As a result, a  quintessence model with a varying equation of state could
display different large scale structure from a $\Lambda$CDM model.
Varying the  equation of state will result in different amounts of dark energy at different times. It has been shown that
models
with a larger density of dark energy at high redshift than $\Lambda$CDM  have more developed large scale structure at early times, when
normalised to the same $\sigma_8$ today \citep{Grossi:2008xh,Francis:2008md}.
In Section \ref{wmap}, we present the simulation results for each quintessence model
where the initial conditions  were generated using a $\Lambda$CDM linear theory power spectrum and
the background cosmological parameters are the best fit values assuming a $\Lambda$CDM cosmology \citep{Sanchez:2009jq}.
The difference
between the simulations is the result of having a  different linear growth rate for the dark matter perturbations.

The presence of small but appreciable amounts of dark energy at early times also modifies the growth
rate of fluctuations from that expected in a matter dominated universe and hence changes the shape of the linear theory $P(k)$ from the $\Lambda$CDM prediction.
As most of the
quintessence models we  consider display a non-negligible contribution to the overall density from dark energy
at early times,
 the matter power spectrum is  affected in two ways \citep{Ferreira:1997hj,Caldwell:2003vp,Doran:2007ep}.
The growth of  modes
on scales $k >k_{\tiny \mbox{eq}}$, where $k_{\tiny \mbox{eq}}$ is the
wavenumber corresponding to the horizon scale at matter-radiation equality, is  suppressed relative to the growth expected in a $\Lambda$CDM universe.
For fluctuations with wavenumbers  $k<k_{\tiny \mbox{eq}}$ during the matter dominated epoch,
the suppression takes place after the mode enters the horizon and
the growing mode is reduced relative to a model with $\Omega_{\tiny \mbox{DE}} \simeq 0$.
The overall result is a scale independent suppression for
subhorizon modes, a scale dependent red tilt ($n_s<1$)
for superhorizon modes and an overall broading of the turnover in the
power spectrum.
This change in the  shape of the turnover in the matter power spectrum can be clearly seen in Fig. \ref{rawpk} for the AS model.
This damping of the growth after horizon crossing will result in a smaller $\sigma_8$ value for the quintessence models compared to $\Lambda$CDM if
normalised to CMB fluctuations (see also \citet{2004PhRvD..70d1301K}).

In Section \ref{bao}, we have used the publicly available PPF (Parametrized Post-Friedmann) module for CAMB, \citep{Fang:2008sn}, to generate the linear theory power spectrum.
Fig. \ref{rawpk} shows the dark matter power spectra at $z=0$ generated by CAMB for each quintessence model and $\Lambda$CDM with the same cosmological
parameters, an initial scalar amplitude of $A_s = 2.14 \times 10^{-9}$
and a spectral index $n_s = 0.96$ \citep{Sanchez:2009jq}. As can be seen in this plot,
models with higher fractional energy densities at early times have a lower $\sigma_8$ today and
a broader turnover in $P(k)$.

Finally, quintessence dark energy models will not necessarily
agree with observational data when adopting
the cosmological parameters derived assuming a $\Lambda$CDM
cosmology.
We use the best fit cosmological parameters for each quintessence model from \citet{Jennings} which fit the
the  observational
 constraints on distances such as the measurements of the angular diameter distance and sound horizon at the last scattering surface
from the cosmic microwave background.

\section{Simulation Details}

We  determine the impact of quintessence dark energy 
on the growth of cosmological structures  through a series of large N-body simulations.
These simulations were carried out at the Institute of Computational Cosmology using a memory efficient version of the  TreePM
  code  {\tt Gadget-2}, called {\tt L-Gadget-2} \citep{Springel:2005mi}. As our starting point, we consider a $\Lambda$CDM model with 
the following cosmological parameters:  
$\Omega_{\rm m} = 0.26$,
 $\Omega_{\rm{DE}}=0.74$, $\Omega_{\rm b} = 0.044$,
$h = 0.715$ and a spectral tilt of $n_{\mbox{s}} =0.96$ \citep{Sanchez:2009jq}. 
The  linear theory rms fluctuation 
in spheres of radius 8 $h^{-1}$ Mpc is set to be  $\sigma_8 = 0.8$.

The simulations use $N=646^3 \sim 269 \times 10^6$ particles to represent the dark matter in a  computational box of 
comoving length $1500 h^{-1}$Mpc. 
The particle mass in the simulation is $9.02 \times 10^{11}  h^{-1}
M_{\textrm sun}$ with a mean interparticle separation of
$r \sim 2.3$ $h^{-1}$Mpc.
The linear theory power spectrum used to generate the initial 
conditions was created using the CAMB package of \citet{Lewis:2002ah}.  
The linear theory $P(k)$ is generated for each quintessence 
model using a modified version of CAMB which incorporates 
the influence of dark energy on dark matter clustering at early times. 
 In each model the power spectra at redshift zero have been normalised to have 
$\sigma_8 = 0.8$. 

\begin{figure} 
{\epsfxsize=9.truecm
 \epsfbox[82 360 424 708]{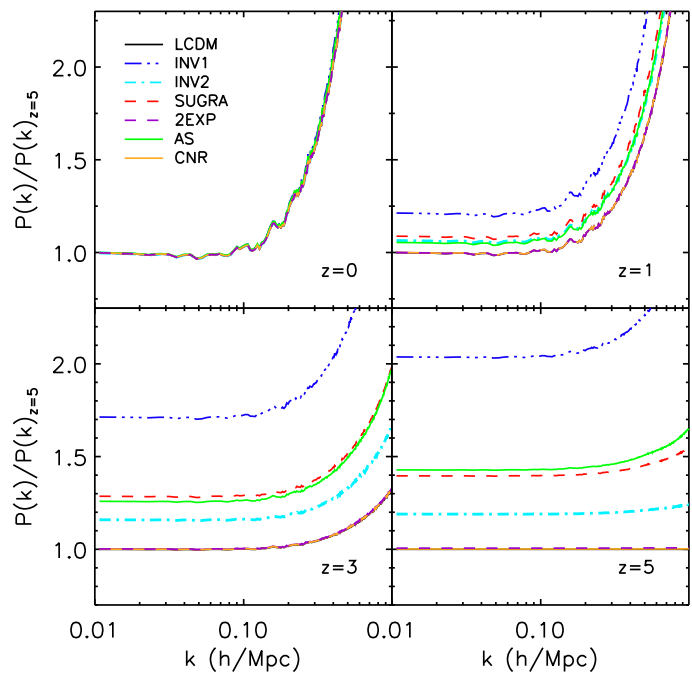}}
\caption{The nonlinear growth of the power spectra in the various quintessence models as indicated by the key in the top left panel.
Each panel shows a different
redshift. The power spectra in each case have been
divided by the $\Lambda$CDM power spectrum at redshift 5 scaled to take out the difference between the $\Lambda$CDM growth factor at $z=5$ and the redshift plotted in the panel.
A deviation of the power ratio from unity therefore indicates a difference in $P(k)$ from the linear perturbation theory of $\Lambda$CDM.}
\label{pk}
 \end{figure}

\section{Results}

\subsection{The matter power spectrum \label{wmap}}

In this first stage of simulations, we investigate the effect on the growth of structure by changing the expansion rate of the universe for each quintessence model. 
The same $\Lambda$CDM initial power spectrum and cosmological parameters were used for all models.

To highlight the differences in the power between the different models, we plot in Fig. \ref{pk} the measured power 
divided by the power at $z=5$, after scaling to take into account the difference in the linear theory growth factors for the output redshift and $z=5$, for $\Lambda$CDM. 
This removes the sampling variance from the plotted ratio \citep{1994MNRAS.270..183B}. 
A ratio of unity  in Fig. \ref{pk} would indicate linear growth at the same rate as expected in $\Lambda$CDM.
\begin{figure}
{\epsfxsize=9.truecm
\epsfbox[71 361 422 703]{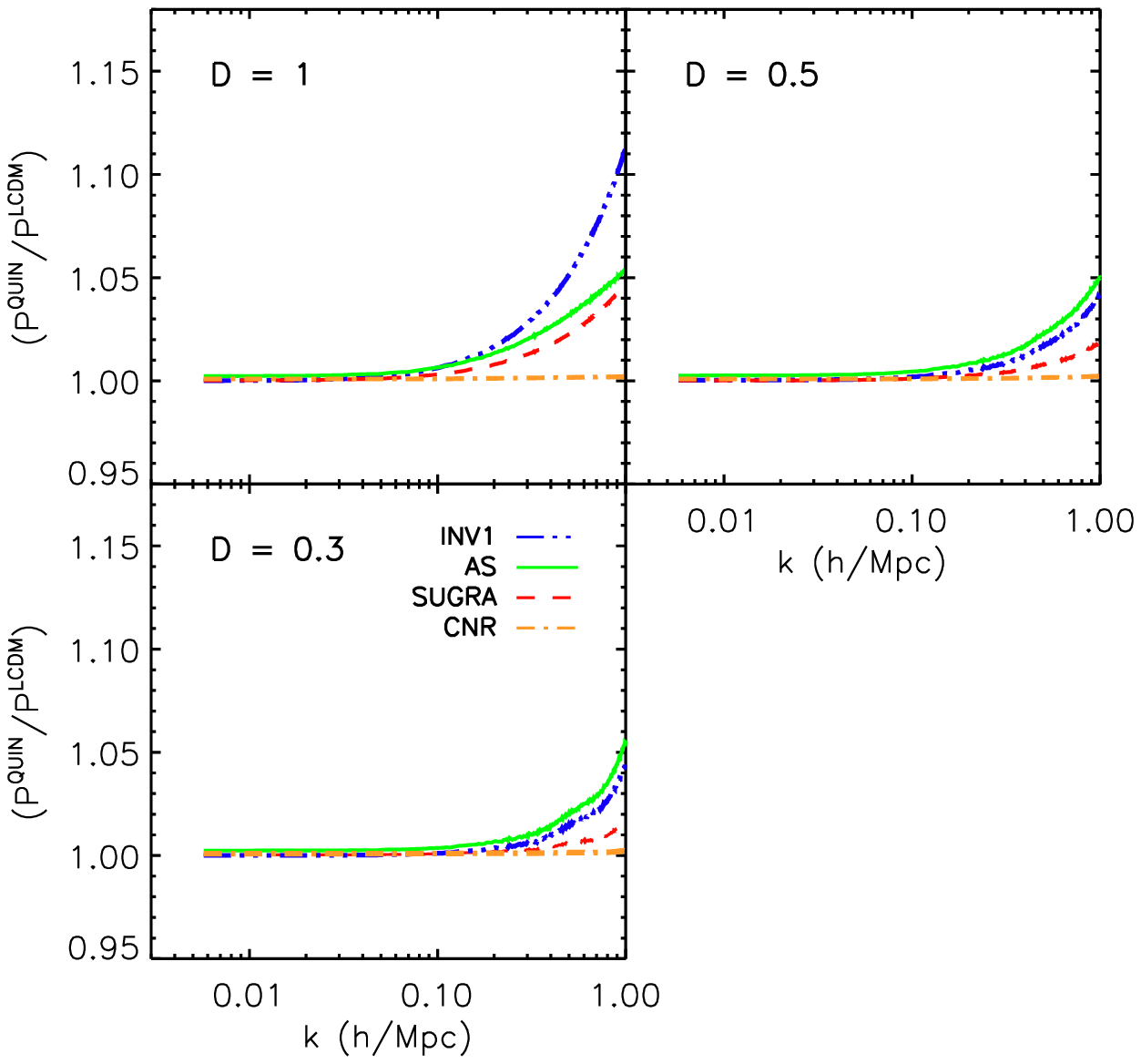}}
 \caption{ The ratio of the quintessence model power spectra to the $\Lambda$CDM
power spectrum output from the simulations at three values of the linear growth factor $D = 1, D = 0.5$
and $D = 0.3$.
Each panel shows the results of this exercise for the AS, CNR, 2EXP and SUGRA quintessence models. The growth factors
correspond to $z=3.4$ ($D=0.3$), $z=1.6$ ($D=0.5$) and $z=0$ ($D=1$) for $\Lambda$CDM.
A ratio of unity would indicate that the growth factor is the only ingredient needed
to predict the power spectrum in the different quintessence models. Note the expanded scale on the y axis.
}
\label{scaledgrowth}
\end{figure}

Fig. \ref{pk} shows four epochs in the evolution of 
the power spectrum for all 
of the quintessence models and $\Lambda$CDM. 
The black line in the plot shows the $P(k)$ ratio for $\Lambda$CDM (note the yellow curve for the 
CNR model is overplotted). 
Non-linear growth can be seen as an increase in the power ratio on small scales, 
$k>0.3h$Mpc$^{-1}$ at $z=3$ and $k>0.1h$Mpc$^{-1}$ at $z=0$.
 Four of the quintessence models (INV1, INV2, SUGRA and AS) differ significantly from $\Lambda$CDM for $z>0$. 
These models  show   advanced
 structure formation i.e. more power than $\Lambda$CDM, and a large increase in the 
 amount of nonlinear growth.
All models are normalised to have $\sigma_8 =  0.8 $ today 
and as a result all the power spectra are very similar at redshift zero in Fig. \ref{pk}.  
The power spectra predicted in the 2EXP and CNR models show minor departures from that in the $\Lambda$CDM cosmology. 
This is expected as Fig. \ref{w}
shows the equations of state  in these two models are the same as $\Lambda$CDM at low redshifts 
and all three simulations began from identical initial conditions.  

Finally, we investigate  if the enhanced growth in the power spectrum seen in Fig. \ref{pk} in the quintessence models is due solely 
to the different linear growth rates at a given redshift in the models. 
In order to test this idea, the power spectrum in a quintessence model and $\Lambda$CDM 
are compared not at the same redshift but at the same linear growth factor \footnote{We thank S. D. M. White for this suggestion.}. 
As the growth rates in some of the 
quintessence models are very different from that in the standard $\Lambda$CDM cosmology, the power spectra
required from the simulation will be at different output redshift in this comparison.
For example, the normalised linear growth factor is $D = 0.5$ at a redshift of $z=1.58$ in a $\Lambda$CDM model and has the same value 
at  $z = 1.82$ in the SUGRA model, at $z = 1.75$ in the AS model and at 
$z=2.25$ in the INV1 quintessence model. 
In Fig. \ref{scaledgrowth} we show the power spectrum of  simulation outputs from the INV1, AS, SUGRA and CNR models divided 
by the power spectrum output in $\Lambda$CDM at the same linear growth rate. We ran the simulations taking three additional redshift
 outputs where the linear growth rate had values of $D = 1, D = 0.5$ and 
$D = 0.3$. 
It is clear from Fig. \ref{scaledgrowth} that scaling the power spectrum in this way can explain the enhanced linear and most of the excess nonlinear growth seen in 
Fig. \ref{pk} for scales $k<0.1 h$Mpc$^{-1}$. 
For example, in the INV1 model the enhanced nonlinear growth,
on scales $k \sim 0.3h$Mpc$^{-1}$ at fixed $D=0.3$, 
differs from $\Lambda$CDM by at most 5\% in Fig. 
\ref{scaledgrowth} as opposed to at most 30\% at $z=5$ in Fig. \ref{pk}.
At earlier redshifts when the linear growth rate is $D=0.3$, the nonlinear growth 
in the quintessence models agrees with
$\Lambda$CDM on  smaller wavenumbers $k<0.3 h$Mpc$^{-1}$. As in Fig. \ref{pk}, the CNR model shows no difference from $\Lambda$CDM when plotted in this way. 

The reason for the success of this simple model - matching the growth factor to predict the clustering - 
can be traced to the universality of the mass function, which is discussed in \citet{Jennings}.
Hence, it seems that scaling the power spectrum using the linear growth rate can be used to predict the linear growth in the quintessence dark energy 
simulations and can reproduce 
some of the nonlinear growth at early 
redshifts. In Fig. \ref{scaledgrowth} there are still some differences in the small scale growth in quintessence models compared to $\Lambda$CDM which cannot be explained
by the different linear growth rates.
We find that nonlinear evolution is not just a function
of the current value of the linear growth rate but also depends on its history through the  evolution of the coupling between long and
short-wavelength modes.

\begin{figure}
{\epsfxsize=10.truecm
\epsfbox[101 370 523 613]{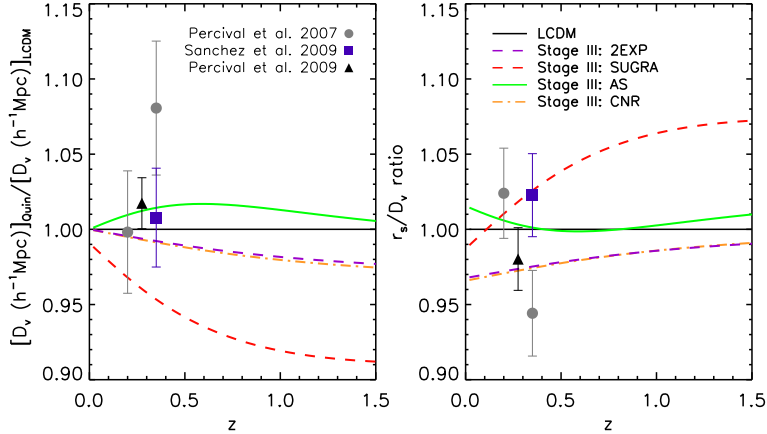}}
 \caption{ The ratio of the distance measure $D_v(z)$ (left panel) and the ratio of $r_s(z_d)/D_v$ (right panel) for four quintessence models compared to $\Lambda$CDM
 as indicated by the key in the right hand panel.
The grey circles are estimates from \citet{Percival:2007yw} at $z=0.2$ and $z=0.35$.
\citet{Sanchez:2009jq} combined CMB data with information on the shape of the redshift space correlation function using a larger
LRG
dataset and found
$D_v(z=0.35) = 1300 \pm 31$ Mpc and $r_s(z_d)/D_v = 0.1185 \pm  0.0032$ at $z=0.35$ (blue squares).
The data points from \citet{Percival:2009xn} for $D_v$ and $r_s(z_d)/D_v$ at $z=0.275$ are plotted as black triangles.
Stage III models match observational distance constraints from CMB, BAO and SNIa measurements.
}\label{RSDV}
\end{figure}

\subsection{The appearance of baryonic acoustic oscillations in quintessence models\label{bao}}

We now examine the baryonic acoustic oscillation signal in the matter power spectrum for the AS, SUGRA and CNR models. 
Each of these simulations uses a consistent linear theory power spectrum  with the best fit cosmological parameters from
 \citet{Jennings}.
\citet{Angulo:2007fw} presented a detailed set of predictions for the appearance of the BAO signal in the $\Lambda$CDM model,
covering the impact of nonlinear growth, peculiar velocities and scale dependent redshift space distortions and galaxy bias. Here we 
focus on the first of these effects and show power spectra in real space for the dark matter, for these selected quintessence models.

The baryonic acoustic oscillations are approximately a standard ruler and depend on the sound horizon, $r_s$ \citep{Sanchez:2008iw}.
 The apparent 
size of the BAO scale depends on the distance to the redshift of observation and on the ratio $r_s/D_v$, where $D_v$ is an effective distance measure 
which is a combination of $D_A$ and $H$.
In most quintessence models, $r_s$ remains unchanged unless there is appreciable dark energy at last scattering. Models which have the same 
ratio of $r_s/D_v$ are impossible to distinguish using BAO.

\begin{figure}
{\epsfxsize=10.truecm
\epsfbox[84 366 536 612]{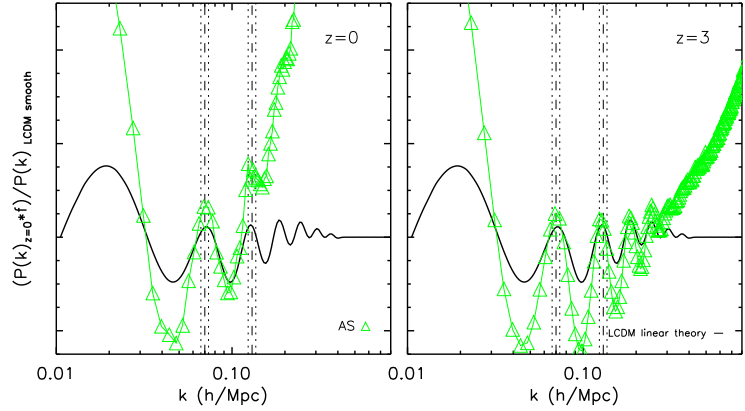}}
 \caption{The real space power spectrum in the AS model (green triangles) on large scales at $z=0$ (left) and
$z=3$ (right). All power spectra have been divided by a smoothed
linear \lq no-wiggle\rq \, theory $P(k)$ for $\Lambda$CDM. The factor, $f$, removes the scatter of the power measured in the simulation
around the expected linear theory power.
  The black solid line represents the linear theory power spectrum in $\Lambda$CDM divided by the smooth reference spectrum.
The vertical dashed (dotted) lines show the position of the first two  acoustic
peaks (positions $\pm 5\%$) for a $\Lambda$CDM cosmology.
}\label{BAOI}
\end{figure}
\begin{figure}
{\epsfxsize=10.truecm
\epsfbox[84 366 536 612]{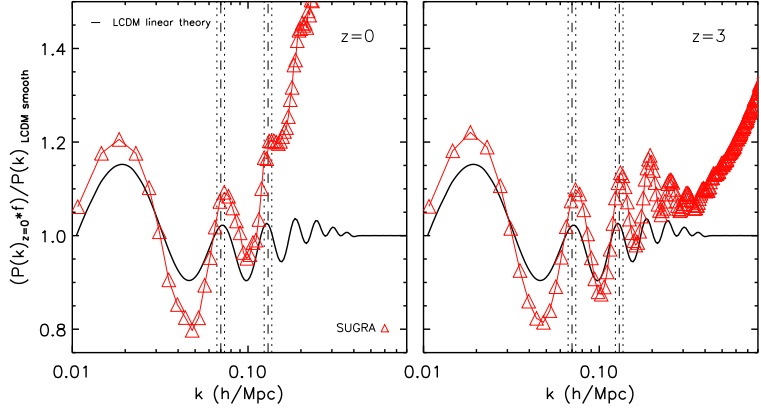}}
 \caption{ The real space power spectrum for the SUGRA model (red triangles) on large scales
at $z=0$ and $z=3$.
}\label{BAOIII}
\end{figure}

To calculate the power spectrum for a galaxy redshift survey, the measured angular and radial separations of galaxy pairs are converted to
co-moving separations. This conversion is dependent on the cosmological model assumed.
These changes can be combined into the single effective distance measure, $D_v$.
Once the power spectrum is calculated in one model we can simply re-scale $P(k)$ using $D_v$ to obtain the
power spectrum and BAO peak positions in another cosmological model (see \citet{Sanchez:2009jq}).
In the left panel of Fig. \ref{RSDV}, we plot the ratio of $D_v$ in four quintessence models compared to $\Lambda$CDM up to $z=1.5$.
\citet{Percival:2007yw} found $D_v = 564 \pm 23 h^{-1}$Mpc at $z=0.2$ and $D_v = 1019 \pm 42 h^{-1}$Mpc at $z=0.35$ using
the observed scale of BAO measured from the SDSS DR5 galaxy sample and 2dFGRS.
These data points are plotted as grey circles in Fig. \ref{RSDV}. 
These authors reported a 2.4$\sigma$ discrepancy between their results using BAO and the constraints available at the time from supernovae. 
The blue square plotted in the left panel in  Fig. \ref{RSDV} is
the constraint  $D_v = 1300 \pm 31$ Mpc at $z=0.35$ found by \citet{Sanchez:2009jq}. This constraint was found using a larger
LRG
dataset and improved modelling of the correlation function on large scales. The constraint found by \citet{Sanchez:2009jq}
using CMB  and BAO data is fully consistent with CMB and SN results.
The results from \citet{Percival:2009xn} for $D_v$ and $r_s(z_d)/D_v$ at $z=0.275$ using WMAP 5 year data together with the SDSS data release 7 galaxy sample are also plotted
(black triangles). The new \citet{Percival:2009xn} results are in much better agreement with those of \citet{Sanchez:2009jq}.

Over the range of redshifts plotted in Fig. \ref{RSDV}, the distance measure, $D_v$, in the AS, 2EXP and CNR models differ from $\Lambda$CDM by at most 2\% and is $<1\%$ in these
models for $z<0.2$. Re-scaling the power spectrum for these dark energy cosmologies
 would result in a small shift $\sim 1\%$ in the position of the BAO peaks at low redshifts.
The value of $D_v$ in the SUGRA model differs from $\Lambda$CDM by at most 9\% up to $z=1.5$.
The right panel in Fig. \ref{RSDV} shows the ratio of $r_s(z_d)/D_v$ in the quintessence models compared to $\Lambda$CDM, where $r_s$ is the
co-moving sound horizon scale
at the drag redshift, $z_d$.
The value of $r_s(z_d)/D_v$ can be constrained using the position of the BAO in the power spectrum. In the right panel of Fig. \ref{RSDV} the grey symbols are the results from
\citet{Percival:2007yw} at $z=0.2$ and $z=0.35$. From this plot it is clear that the SUGRA and AS model are within the 1$\sigma$ limits at $z=0.2$.
The 2EXP and CNR model lie just outside the 1$\sigma$ errors at $z=0.35$. Note the value of $r_s(z_d)/D_v$ for $\Lambda$CDM at $z=0.35$ also lies
outside the  1$\sigma$ errors (see \citet{Percival:2009xn} for more detail). The blue square plotted in the right panel in 
Fig. \ref{RSDV} is $r_s(z_d)/D_v= 0.1185 \pm  0.0032$ at $z=0.35$ and was obtained using  
information on the redshift space correlation function together with CMB data \citep{Sanchez:2009jq}.

In Fig. \ref{BAOI} and \ref{BAOIII}
 we plot the $z=0$ and $z=3$ power spectra in the AS and SUGRA models divided 
by a linear theory $\Lambda$CDM reference spectrum  which has been smoothed using the coarse rebinning method proposed by
 \citet{Percival:2007yw} and refined by \citet{Angulo:2007fw}. After dividing 
by this  smoothed  power spectrum, the acoustic peaks are more visible in the quasi-linear regime.
In Figs. \ref{BAOI} and \ref{BAOIII}, the 
measured power in each bin has been multiplied by a factor, $f$, to  
remove
the scatter due to the small number of large scale modes in the simulation \citep{1994MNRAS.270..183B,Springel:2005nw}. 
This factor, $f = P(k)_{\tiny \mbox{linear}}/P(k)_{\tiny \mbox{N-body}}$, 
is the ratio of the expected linear theory power and the measured power in each bin at $z=5$,
at which time  the power on these scales is still in the 
linear regime.

In Fig. \ref{BAOI}  we plot the AS power spectrum  as  
green triangles . The black line represents
 the linear theory power in $\Lambda$CDM divided by the smooth reference spectrum.
In both plots and for all power spectra, the same reference spectrum is used. 
 The difference between the 
AS and $\Lambda$CDM linear theory, as shown in Fig. \ref{rawpk}, results in an increase 
in large scale power on scales $k<0.04 h$Mpc$^{-1}$.
The vertical dashed (dotted) lines show the positions of the first two acoustic peaks 
(positions $\pm 5\%$) for a $\Lambda$CDM cosmology.

When the best fit  cosmological parameters and a consistent linear theory power spectrum are used for the AS model, the sound 
horizon in the AS model and  $\Lambda$CDM are very similar at $z \sim 1090$ and there is a very small ($<1$\%)
 shift in the position of the first peak (green triangles). As 
there is less nonlinear growth at $z=3$ the 
 higher order peaks are more visible in the right-hand panel in Fig. \ref{BAOI}.

In Fig. \ref{BAOIII}, the SUGRA power spectrum is plotted using consistent cosmological parameters and linear theory power spectrum.
We find a shift of $\sim 5\%$ in the position of the
  first peak in the SUGRA model compared to  $\Lambda$CDM.
Note the units on the x axis are $h/$Mpc and, $h=0.67$ for the 
SUGRA model  compared to $h=0.715$ for $\Lambda$CDM \citep{Jennings}.
On small scales 
the BAO signature is damped due to more nonlinear  structure formation at $z=0$ compared to $z=3$ as shown in Fig \ref{BAOIII}. 
We find a large increase in the power in the region of the
 second peak, $k \sim 0.15 h$Mpc$^{-1}$ in both the AS and SUGRA models compared to $\Lambda$CDM.
For brevity we have not included the plots of the power spectra for the CNR model showing the baryonic acoustic oscillations. We find identical peak positions 
in $\Lambda$CDM and this model  at $z=0$.

The AS and SUGRA model are very different to $\Lambda$CDM at late times and as result they affect the growth of structure at $z>0$ as seen in Section \ref{wmap}. 
 We have found that models like this do not necessarily have  different BAO peak positions to $\Lambda$CDM in the matter power spectrum.
These results  suggest that 
distinguishing a quintessence model, like the AS  model considered here, using measurements of the BAO peak positions in future 
galaxy surverys, will be extremely difficult. The BAO peak positions for the  CNR model will be shifted by at most 2\% in the range $z<1.5$
compared to $\Lambda$CDM after re-scaling the power spectra by $D_v$. In conclusion it is possible to have  
quintessence cosmologies with  higher levels of dark energy at early times than in $\Lambda$CDM, 
but which predict the same peak positions for the BAO in the matter power spectrum.

\section{Conclusions and Summary}
 
Observing the dynamics of dark energy is the central goal of future wide-angle galaxy surveys and would distinguish a cosmological constant from a dynamical quintessence model.
We have analysed the influence of dynamical dark energy on structure formation using N-body simulations with
a range of quintessence models.
The majority of the
models could be classified as \lq early dark energy\rq \, models as they have a non-negligible amount of
dark energy at early times \citep{Jennings}.

In order to accurately mimic the dynamics of the original quintessence 
models at high and low redshift, it is necessary to use a general 
prescription for the dark energy equation of state 
which has more parameters than the popular 2 variable equation. Parametrisations for $w$ which use
 2 variables are unable to faithfully 
represent dynamical dark energy models  over a wide range of redshifts, certainly not sufficient to run an N-body simulation,  
and can lead to biases when used to constrain parameters \citep{Bassett:2004wz}. 
We use the 
parametrization of \citet{Corasaniti:2002vg} which accurately describes the dynamics of the different quintessence models.
With this description of the equation of state, our simulations are able to accurately describe the impact of the quintessence model on the expansion 
rate of the universe, from the starting redshift to the present day. This would not be the case with a 2 parameter model for the equation of state.

We have taken into account three levels of modification to a
$\Lambda$CDM cosmology in order to faithfully incorporate 
the effects of quintessence dark energy into a N-body simulation.
The first step is to replace the cosmological constant with a quintessence model with an
equation of state different from $w=-1$ which will lead to a universe with a different expansion history. 
The second step is to allow the change in the expansion history and perturbations in the quintessence field to have an impact on the form of the 
linear theory power spectrum.
 The  shape of the power spectrum can differ significantly from 
$\Lambda$CDM on large scales if there is a non-negligible 
amount of dark energy present at early times.
Thirdly, as the quintessence model should be consistent with observational constraints, 
the cosmological parameters used for the dark energy model could be different from the 
best fit $\Lambda$CDM 
parameters.

In the first stage of comparison, in which all that is changed is the expansion history of the universe, we found that some of the quintessence models showed
enhanced structure formation at $z>0$ compared to $\Lambda$CDM.
The  INV1, INV2, SUGRA and AS models  have slower 
growth rates than $\Lambda$CDM. Hence, when normalising to the same $\sigma_8$ today, 
structures must form at earlier times in these models to 
overcome the lack of  growth at late times.
The difference in linear and nonlinear growth can largely be explained by the difference in the growth factor at different epochs in the models.
At the same growth factor, the power in the models diverges at the 15\% level well into the nonlinear regime.

We will now summarise and discuss the main results for each model.
The full results for the matter power spectrum and mass function for each model can be found  in \citet{Jennings}.
As found in \citet{Jennings}, the INV1 model was unable to fit the data with a reasonable $\chi^2/\nu$ (Table A3). 
This model has the largest growth factor ratio to $\Lambda$CDM at $z=5$ and 
as a result showed the most enhanced growth in our simulations.
In the 2EXP model, the rapid transition to $w=-1$ in the equation of state early on leaves little impact on the growth of dark matter and as a result the
power spectra and mass function are indistinguishable from $\Lambda$CDM.

The SUGRA model has   
enhanced linear and nonlinear growth and halo abundances compared to 
$\Lambda$CDM at $z>0$ and an altered linear theory power spectrum shape. 
Analysing the SUGRA power spectra,
 from a simulation which used the best fit parameters for this model, reveals a $\sim 5\%$ shift in the position of the first BAO 
peak. We find the distance measure $D_v$ for the SUGRA model 
differs by up to 9\% compared to $\Lambda$CDM  over the range $0<z<1.5$. Re-scaling the power measured for the SUGRA model by the difference in $D_v$ 
would result in an even larger shift in the position of the 
BAO peaks.

The CNR model has high levels of dark energy early on
which alters the spectral shape on large scales.
This model has BAO peak positions at $z=0$ which are the same as in $\Lambda$CDM. 
For $z<0.5$ the distance measure, $D_v$, for the CNR model 
differs from  $\Lambda$CDM by $\sim 1\%$, as result there would be a corresponding small shift in the BAO peak positions. 
The rapid early transition at $z=5.5$ in the equation of state to $w=-1$ in this model seems to remove any signal of the large amounts of 
dark energy at early times which alters the growth of dark matter perturbations at high redshifts.

The AS model has the highest levels of dark energy at early times, and  
this results in a large increase in large scale power, when we normalise the power spectrum to $\sigma_8 =0.8$ today. 
The results using the best fit parameters show both enhanced linear and nonlinear 
growth at $z<5$. The linear theory $P(k)$ is altered on scales $k\sim 0.1 h$Mpc$^{-1}$ which 
drives an increase in nonlinear growth on small scales compared to $\Lambda$CDM. 
We find that the  AS model produces a BAO profile
 with peak positions similar to those in $\Lambda$CDM. At low redshifts there is an  $\sim 1\%$ shift in the first peak compared 
to $\Lambda$CDM  after re-scaling the power with the difference in the distance measure $D_v$ between the two cosmologies.

These results from our N-body simulations 
show that dynamical dark energy models in which the dark energy equation of state makes a late $(z<2)$ rapid transition 
to $w=-1$ show enhanced linear and nonlinear growth compared to  $\Lambda$CDM at $z>0$. 
We found that dynamical dark energy 
models can be significantly different from $\Lambda$CDM at late times and still produce similar BAO peak positions in the matter power spectrum.
Models which have a rapid early transition in their dark energy equation of state and mimic $\Lambda$CDM after the transition, show the same linear and nonlinear growth 
 as $\Lambda$CDM for all redshifts. We have found that these models can give rise to BAO peak positions
in the matter power spectrum which are the same as those
 in a $\Lambda$CDM cosmology. This is true despite these models having non-negligible amounts of dark energy at early times.

Overall, our analysis shows  
parameter degeneracies allow some quintessence models 
to have identical BAO peak positions to $\Lambda$CDM and so these measurements alone will not be able 
to rule out some quintessence models.
Although including the dark energy
perturbations has been found to increase 
these degeneracies \citep{Weller:2003hw}, incorporating them into the N-body code  would clearly be 
the next step towards simulating quintessential dark matter 
with a  fully physical model.

\begin{theacknowledgments}
EJ acknowledges receipt of a fellowship funded by the European
 Commission's Framework Programme 6, through the Marie Curie Early Stage
Training project MEST-CT-2005-021074.
This work was supported in
part by grants from the U.K. Science and Technology Facilities
Council held by the Extragalactic Cosmology Research Group and 
the Institute for Particle Physics Phenomenology at Durham University.
We acknowledge helpful conversations with Simon D. M. White, Ariel G. S\'{a}nchez, Shaun Cole and Lydia Heck for support running the simulations.
\end{theacknowledgments}

\bibliographystyle{aipproc}
\bibliography{mybibliography}

\label{lastpage}

\end{document}